\documentstyle[12pt]{article}
\let\section=\subsection     \let\subsection=\subsubsection                
\def\ud{\stackrel{>}{<}}
\def\du{\stackrel{<}{>}}
\input prepictex
\input pictex
\input postpictex

\begin{document}
\begin{center}
   {\large \bf DEVELOPING TRANSPORT THEORY TO }\\[2mm]
   {\large \bf SYSTEMATICALLY INCLUDE MESONS } \\[2mm]
   {\large \bf AND HADRONIZATION}\\[5mm]
   S.P.~KLEVANSKY, P.~REHBERG, A.~OGURA and J.~H\"UFNER  \\[5mm]
   {\small \it  Institut f\"ur Theoretische Physik \\
   Philosophenweg 19, D-69120 Heidelberg, Germany \\[8mm] }
\end{center}

\begin{abstract}\noindent
Transport theory for the Nambu--Jona-Lasinio (NJL) model is examined.   The
collision term is investigated in a  first approach
 in a coupling strength expansion.   It
can
be demonstrated that this leads to a form easily recognized from the
Boltzmann equation if the quasiparticle approximation is used.    It is seen
that enforcing a quasiparticle approximation suppresses the three body
creation and annihilation processes that would otherwise be present.
Including mesons and hadronization consistently in a $1/N_c$ expansion is
discussed briefly.    Some comments on the numerical simulation of the
Vlasov equation and the relaxation time approximation to the collision term
are made.
\end{abstract}

\section{Introduction}
It is a well accepted fact that the property of chiral symmetry is essential
for a correct description of the static properties of the low lying mesons.
Further, it is known from lattice simulations that, at a finite value of the
temperature, this symmetry is restored.   This, in turn, may lead to
substantial
changes in the mesonic spectrum, something that we know only from model
calculations \cite{reh1}.  Now, with the increasing interest in heavy-ion
collisions,
and in the non-equilibrium descriptions thereof, this particular feature of
quantum chromodynamics (QCD) has to date been ignored.   Thus, the purpose
of this work is to construct a transport theory that is based on a model of
QCD that contains the essential feature of chiral symmetry breaking and
restoration, and to investigate by simulation whether {\it dynamical}
effects
due to this symmetry breaking / restoration are evidenced that could have
experimental consequences.

Our starting point is the Nambu--Jona-Lasinio (NJL) model, that
intrinsically
contains only quark degrees of freedom that interact via an effective two
and three body interaction for three flavors.   Mesonic states are
constructed as collective excitations in this model.    No explicit gluonic
degrees of freedom are included.    A transport theory for this Lagrange
density,
at the lowest level, involves only the quark degrees of freedom.  One is
required to go to higher orders -- in this strong coupling theory, an
expansion in the inverse number of colors $1/N_c$ is appropriate --
in order to include the mesonic degrees of freedom.   Doing so, one finds
that it becomes possible to construct a consistent theory of quarks and
mesons in which the collision term contains {\it inter alia} the
hadronization processes $q\bar q'\rightarrow MM'$ in addition to the usual
elastic scattering $q\bar q'\rightarrow q\bar q'$ and $qq'\rightarrow qq'$
in the quark sector.   In this talk, we sketch the development of the
collision term first expanded in the interaction strength, and indicate how
mesons are included.   Enforcing a quasiparticle approximation is seen to
suppress the three body creation and annihilation processes that would
otherwise be present.   The expansion shows how meson dynamics
are incorporated naturally into the one-body transport equation for the
quark or antiquark distribution function.

Finally, we indicate briefly the current status of the numerical
simulations.

\section{Including the collision term in transport theory}

\subsection{Equations of motion}

Non-equilibrium phenomena are completely desccribed via the
Schwinger - Keldysh formalism for Green functions \cite{schw}.
   Many conventions
exist in the literature.   For our purposes, it is simplest to use
the convention of Landau \cite{landau}.   In this, the designations $+$ and $-$
are
attributed to the closed time path that is shown in Fig.1, and the
fermionic Green functions are defined as
\begin{eqnarray}
iS^c(x,y) &=& \left <T\psi(x) \bar \psi(y)\right > = iS^{--}(x,y)\nonumber \\
iS^a(x,y) &=& \left <\tilde T\psi(x)\bar\psi(y)\right > = iS^{++}(x,y)
\nonumber \\
iS^{>}(x,y) &=& \left <\psi(x)\bar\psi(y)\right > = iS^{+-}(x,y) \nonumber \\
iS^<(x,y) &=& -\left <\bar\psi(y)\psi(x)\right > = iS^{-+}(x,y).
\end{eqnarray}
In a standard fashion, one
constructs equations of motion for the matrix of Green functions and one
moves to relative and centre of mass variables, $u=x-y$ and $X=(x+y)/2$.  A
Fourier transform with respect to the relative coordinate, or Wigner
transform, \begin{equation} S(X,p) = \int d^4u e^{ip\cdot u} S\left(X+\frac
u2,X-\frac u2\right) \end{equation} is then performed.  Of particular interest
is
the equation of motion for $S^{-+} =S^<$.  Regarding this, together with the
equation of motion for the adjoint function $S^{<\dagger}$, one arrives at
the so-called transport and constraint equations by adding and subtracting
these.  One finds \begin{equation} \frac{i\hbar}2\{\gamma^\mu,\frac{\partial
S^<}{\partial X^\mu}\} + p_\mu[\gamma^\mu,S^<] = \Sigma^<S^>-\Sigma^>S^< +
[\Sigma^A,S^>] + [\Sigma^R,\Sigma^>], \label{full1} \end{equation} and
\begin{equation} \frac{i\hbar}2[\gamma^\mu,\frac{ \partial S^<}{\partial
X^\mu}] + \{ \not p - m_0,S^<\} = \Sigma^<S^> -\Sigma^>S^< +
\{\sigma^A,S^>\} - \{S^R,\Sigma^>\} \label{full2} \end{equation} which are
general equations that are generic for {\it any} fermionic theory
\cite{heinz}.

\begin{center}

\setcoordinatesystem units <1cm, 1cm> point at 0 0
\unitlength1cm \large

\beginpicture

\arrow <3mm> [0.25,0.75] from 0 1 to 12 1

\arrow <5mm> [0.25,0.75] from 1 1.5 to 6 1.5
\plot 6 1.5 11 1.5 /
\circulararc -180 degrees from 11 1.5 center at 11 1
\arrow <5mm> [0.25,0.75] from 11 0.5 to 5.5 0.5
\plot 5.5 0.5 1 0.5 /
\put{$-$} [c] at 5.75 2
\put{$+$} [c] at 5.75 0

\put{ } [l] at 0 -0.5
\put{ } [l] at 12 2.0

\endpicture

\begin{minipage}{13cm} \baselineskip=12pt
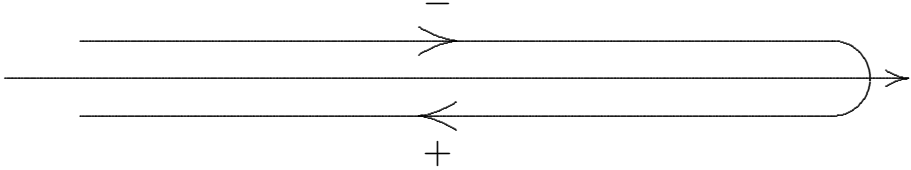
{\begin{small} Fig.~1. Closed time path on which Green functions are
defined. \end{small}} \end{minipage} \end{center}

\subsection{Methods of solution}
A complete method of solution of Eqs.(\ref{full1}) and (\ref{full2})
 follows on inserting the spinor decomposition
\begin{equation}
S^< = F + i\gamma_5 P + \gamma^\mu V_\mu + \gamma^\mu\gamma_5 A_\mu + \frac
12\sigma^{\mu\nu}S_{\mu\nu},
\end{equation}
for $S^<$ into the above equation.    Unfortunately, this is only practical
for the collisionless system \cite{woitek}, and it leads to a set of $16 \times
2$
equations that must be solved simultaneously.   In order to understand the
collision term, it is more useful to insert the Ansatz
\begin{eqnarray}
S^<(X,p) &=& 2\pi i \frac 1{2E_p}[\delta(p_0 -E_p)\sum_{s,s'}u_{s'}(p)
\bar u_s(p) f^{ss'}_q(p,X) \nonumber \\
 &+& \delta(p_0+E_p) \sum_{ss'} v_{s'}(-p)\bar
v_s(-p)\bar f_{\bar q}^{ss'}(-p,X)],
\label{decomp}
\end{eqnarray}
with $\bar f = 1-f$, and evaluate the determining equation for the quark and
antiquark distribution functions $f_q$ and $f_{\bar q}$ (note that
$f_q^{ss'} = \delta_{ss'}f_q$).

\subsection{The collision term}

It is useful to understand the simple collision graphs that are given
in Fig.~2,
where the interaction is displayed as extended for convenience.
\begin{center}
\setcoordinatesystem units <1cm, 1cm> point at 0 0
\unitlength1cm \large

\beginpicture

\plot 4 1 7.5 1 /
\circulararc -88 degrees from 4.5 2 center at 5.75 0.6875
\circulararc  88 degrees from 4.5 2 center at 5.75 3.31

\plot 9.5 1 12.5 1 10 2.5 13 2.5 /
\put{$-i\Sigma^{\rm coll}_{-+} =$} [r] at 3.5 1.5
\put{$+$} [c] at 8.5 1.5

\normalsize
\put{$+$} [c] at 7 0.5
\put{$+$} [c] at 7.5 2
\put{$-$} [c] at 4.5 0.5
\put{$-$} [c] at 4 2

\put{$-$} [c] at 10 0.5
\put{$-$} [c] at 9.5 2.5
\put{$+$} [c] at 13 1
\put{$+$} [c] at 12.5 3

\setdashes
\plot 4.5 1 4.5 2 /
\plot 7 1 7 2 /
\plot 10 1 10 2.5 /
\plot 12.5 1 12.5 2.5 /

\put{ } [l] at 1.5 0
\put{ } [l] at 12 3

\endpicture

\begin{minipage}{13cm}
\baselineskip=12pt
{\begin{small}
Fig.~2. Lowest order contributions to the collision term. \end{small}}
\end{minipage}
\end{center}
Then the terms on the right hand side of Eq.(\ref{full1})
that describe collisions are
\begin{equation}
C.I. = \Sigma^>(X,p)S^<(X,p) - \Sigma^<(X,p)S^>(X,p),
\end{equation}
and which represent loss and gain terms respectively.  For the simplest
form of the NJL model \cite{woitek}, three possible forms for $\Sigma$ exist,
$\Sigma_\sigma$, $\Sigma_\pi$ and $\Sigma_{mixed}$, which contain, for the
spinor parts of the interaction vertices, unity, $i\gamma_5$ throughout, or
$i\gamma_5$ and unity in a mixed fashion \cite{ogu}.   For example,
\begin{eqnarray}
& &\Sigma^{\ud}_\sigma(X,p) =
-4G^2\int\frac{d^4p_1}{(2\pi)^4}\int\frac{d^4p_2}{(2
\pi)^4}\int \frac{d^4p_3}{(2\pi)^4} (2\pi)^4 \delta(p-p_1+p_2-p_3) \nonumber \\
& & \hspace{1.0cm}
\times[S^{\ud}(x,p_1) {\rm{tr}}(S^{\du}(X,p_2)S^{\ud}(X,p_2)) -
S^{\ud}(X,p_1)S^{\du}(X,p_2)S^{\ud}(X,p_3)]. \nonumber \\
\end{eqnarray}
Now the insertion of Ansatz (\ref{decomp}) can be used.   One finds that
\begin{itemize}
\item the combination of all three types of self-energies is required in
order to construct the complete relativistic scattering matrix elements
for quark-quark and quark-antiquark scattering, and
\item processes that lead to the production/destruction of quark-antiquark
pairs are present even at this level \cite{ogu}, but are zero if on-shellness
is
stipulated.
\end{itemize}

  The loss term, obtained as
\begin{equation}
{\rm{tr}}\int_{\Delta_+} dp_0(C.I.)_{\rm{loss}}
\end{equation}
where $\Delta_+$ is a restriction over positive energies is then found to be
\begin{eqnarray}
I_q^{loss} &=&
-\frac\pi{E_p}\int\frac{d^3p_1}{(2\pi)^32E_{p_1}}\frac{d^3p_2}
{(2\pi)^32E_{p_2}}\frac{d^3p_3}{(2\pi)^32E_{p_3}} (2\pi)^4\delta^4(p-p_1+p_2
-p_3)  \nonumber \\
&\times& \Big\{
\frac 12\sum|M_{qq\rightarrow qq}(p2\rightarrow 13)|^2\bar f_q(p_1)
f_q(p_2)\bar f_q(p_3) f_q(p) \nonumber \\
&+& \sum |M_{q\bar q\rightarrow q\bar q}
(p2\rightarrow 1 3)|^2\bar
f_q(p_1)\bar f_{\bar q}(p_3) f_{\bar q}(p_2) f_q(p) \Big\},
\end{eqnarray}
where $M(p2\rightarrow 1 3 ) = M^\sigma(p2\rightarrow 1 3) +
M^\pi(p2\rightarrow 13)$.       Thus one has the relativistic generalization
of the nonrelativistic calculation of Kadanoff and Baym \cite{kaba}.    At this
level,
only elastic scattering is incorporated.

\subsection{Higher orders in $1/N_c$ and meson production}

Although the previous calculation is informative, an expansion in the
coupling strength is inadmissable since $G\Lambda^2\sim 2$.   We thus
briefly delineate the extension to higher order, leaving the details to a
forthcoming publication \cite{sandi}.    To do this, we note that the
expansion in $1/N_c$ must be performed in such a way as to be symmetry
conserving.   The next order terms have been investigated in detail in
Ref.\cite{lemmer}.   The self-energy, in the next order, contains meson
exchange, leading to a self-energy of the form
\begin{eqnarray}
\Sigma(k) &=& 4iGN_c\int \frac{d^4p}{(2\pi)^4} \rm{tr} S^F(p) \\
&-& \int\frac{d^4q}{(2\pi)^4}\{(-iD_\sigma^F(q))i\gamma_5 S^F(k-q) i\gamma_5
+ (-iD_\sigma^F(q)) S^F(k-q) \} \nonumber,
\end{eqnarray}
where the superscript $F$ describes the self-consistently calculated Green
function at this level, see Fig.~3.
\begin{center}
\setcoordinatesystem units <1cm, 1cm> point at 0 0
\unitlength1cm \large

\beginpicture

\plot 3 0 5 0 /
\plot 7 0 11 0 /
\circulararc 360 degrees from 4 1 center at 4 1.5
\circulararc -180 degrees from 7.5 0 center at 9 0
\circulararc -180 degrees from 7.6 0 center at 9 0

\setdashes
\plot 4 0 4 1 /

\put{$-i\Sigma^F =$} [r] at 2.5 0.5
\put{$+$} [c] at 6 0.5

\normalsize
\put{$F$} [l] at 4.7 1.5
\put{$F$} [l] at 10.5 1

\put{ } [l] at 1.5 0
\put{ } [l] at 12 2.0

\endpicture

\begin{minipage}{13cm}
\baselineskip=12pt
{\begin{small}
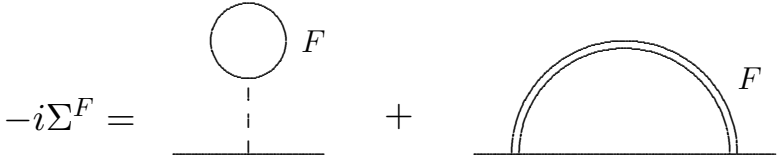
Fig.~3. Self-consistent self energy that includes meson exchange. \end{small}}
\end{minipage}
\end{center}
Expanding these in terms of the Hartree propagator for $S$, and the lowest
order pion irreducible polarization
\begin{equation}
D_\sigma^F(q)^{-1} = 1-2G\Pi^F(p) = 1-2G(\Pi^0+\delta\Pi),
\end{equation}
see Fig.~4, enables one to obtain {\it inter alia}  the  diagrams for $\Sigma$
that are shown in Fig.~5.
\begin{center}

\setcoordinatesystem units <1cm, 1cm> point at 0 0
\unitlength1cm \large

\beginpicture

\circulararc -111 degrees from 3 3.5 center at 4.5 2.4375
\circulararc  111 degrees from 3 3.5 center at 4.5 4.5625

\circulararc  180 degrees from 9.3 4.15 center at 10 4.15
\circulararc  180 degrees from 9.4 4.15 center at 10 4.15
\circulararc -111 degrees from 8.5 3.5 center at 10 2.4375
\circulararc  111 degrees from 8.5 3.5 center at 10 4.5625

\circulararc -111 degrees from 3 1 center at 4.5 -0.0625
\circulararc  111 degrees from 3 1 center at 4.5  2.0625
\plot 4.45 0.2 4.45 1.8 /
\plot 4.55 0.2 4.55 1.8 /

\plot 8 1 9 2 9 0 8 1 /
\plot 12 1 11 2 11 0 12 1 /
\plot 9 2 11 2 /
\plot 9 1.9 11 1.9 /
\plot 9 0 11 0 /
\plot 9 0.1 11 0.1 /

\put{$+$} [c] at 7 3.5
\put{$+$} [c] at 7 1
\put{$+$} [c] at 2 1
\put{$-i\Pi^F=$} [r] at 2.2 3.5

\put{ } [l] at 0.5 0
\put{ } [l] at 12 4.5

\endpicture
\begin{minipage}{13cm}
\baselineskip=12pt
{\begin{small}
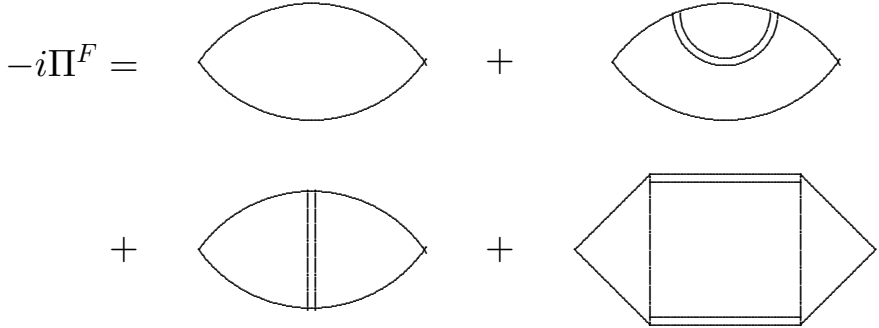
Fig.~4. Contributions to the irreducible polarization to next order in
$1/N_c$.
\end{small}}
\end{minipage}
\end{center}
Intuitively it is obvious that cutting rules applied to these graphs (the
first one, we term ``rainbow'') lead to
the hadronization cross sections as were calculated in Ref.\cite{reh1}
and which now form part of the collision integral.   Thus they are incorporated
naturally into
the transport formalism and are in fact essential in a consistent $1/N_c$
expansion.     Details of this calculation will appear in a forthcoming
publication \cite{sandi}.
\begin{center}
\setcoordinatesystem units <1cm, 1cm> point at 0 0
\unitlength1cm \large

\beginpicture

\plot 0.5 0 5.5 0 /
\plot 7.5 0 13.5 0 /

\circulararc -180 degrees from 1 0 center at 3 0
\circulararc -180 degrees from 1.1 0 center at 3 0
\circulararc -180 degrees from 2 0 center at 3 0
\circulararc -180 degrees from 2.1 0 center at 3 0

\circulararc 360 degrees from 9 1.5 center at 9.5 1.5
\circulararc 360 degrees from 12 1.5 center at 11.5 1.5
\circulararc -67.38 degrees from 8 0 center at 9.626 0
\circulararc -67.38 degrees from 8.1 0 center at 9.626 0
\circulararc  67.38 degrees from 13 0 center at 11.375 0
\circulararc  67.38 degrees from 12.9 0 center at 11.375 0

\plot 9.5 2 11.5 2 /
\plot 9.8 1.9 11.2 1.9 /
\plot 9.5 1 11.5 1 /
\plot 9.8 1.1 11.2 1.1 /

\put{$+$} [c] at 6.5 1

\put{ } [l] at 0.5 -0.2
\put{ } [l] at 12 2.5

\endpicture

\begin{minipage}{13cm}
\baselineskip=12pt
{\begin{small}
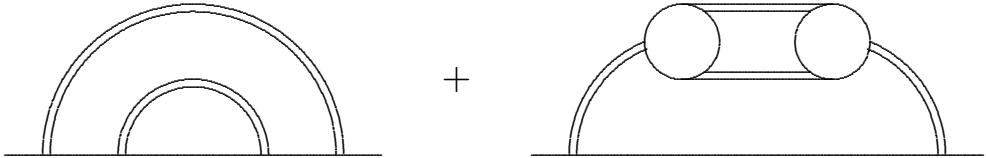
Fig.~5. Diagrams occurring in $\Sigma$ that when cut, lead to hadronization
of a quark and antiquark into two mesons.
\end{small}}
\end{minipage}
\end{center}

\section{Numerical simulations}
Assuming a radial symmetry, the Vlasov equation for the NJL model
\cite{woitek,wilets} has been directly integrated using a finite difference
method \cite{peter}.  Starting with a temperature of $T=240$MeV, and $\mu=
200$MeV and a factorizable profile $f(r,p) = f_r(r)f_p(p)$ with
$f_r(r) = (1-\tanh((r-r_0)/\delta r))/2$, $r_0=3.1$fm and $\delta r = 1$fm
 and $f_p(p) =
2N_cN_f/(\exp(\beta(E-\mu)+1)$,
we find that the symmetry breaking order parameter is globally almost
constant at times not exceeding $17$fm/$c$, setting a maximal time for
the symmetry breaking in this system.

Calculations that include the collision integral
 via a relaxation time approximation are
at present underway.

\section*{Acknowledgments}
This work has been supported in part by the Deutsche Forschungsgemeinschaft
DFG under the contract number Hu 233/4-4, and by the German Ministry for
Education and Technology under contract number 06 HD 742.

\end{document}